\documentclass[twocolumn,english,aps,prb,floatfix,preprintnumbers,showpacs,amsfonts,amssymb]{revtex4}
\usepackage[T1]{fontenc}
\usepackage[latin1]{inputenc}
\usepackage{amsmath}
\usepackage{amssymb}

\makeatletter

\providecommand{\LyX}{L\kern-.1667em\lower.25em\hbox{Y}\kern-.125emX\@}

\usepackage{graphicx}
\usepackage{graphicx}
\usepackage{graphicx}
\usepackage{amscd}
\usepackage{bm}

\usepackage{babel}
\makeatother
\begin{document}
\title{Transport properties of a Luttinger liquid in the presence of several time-dependent impurities}
\newcommand{\be}{\begin{equation}}
\newcommand{\ee}{\end{equation}}
\newcommand{\bd}{\begin{displaymath}}
\newcommand{\ed}{\end{displaymath}}

\author{D.\ Makogon, V.\ Juricic, and C.\ Morais Smith}

\affiliation{Institute for Theoretical Physics, University of
Utrecht, Leuvenlaan 4, 3584 CE Utrecht, The Netherlands.}

\date{\today{}}

\begin{abstract}
We study the transport properties of a Luttinger liquid in the
presence of several time-dependent weak point-like impurities. Our
starting point is the bosonized form of the Luttinger liquid
Hamiltonian with a potential introduced by the impurities. We find
the correction to the total current due to the backscattering of
the electrons by the impurities in first order of perturbation
theory for a general time-dependent impurity potential. We then
apply this result to a single time-dependent impurity, and show
that it reproduces the result obtained by Feldman and Gefen in
Ref.\ \onlinecite{Feldman}, in particular that the current
enhancement occurs only for strong repulsive electron-electron
interactions. For the case of two impurities oscillating with the
same frequency, we find a novel effect, namely, that the
dc-component of the backscattering current is positive {\it even
for weak repulsive interactions}, due to the presence of the
interference term induced by spatial correlations in the Luttinger
liquid. We show that the impurity frequencies at which the
enhancement takes place form a band-like structure. In the case of
several impurities, the bands become broader, because of
interference effects.
\end{abstract}


\maketitle

\section{Introduction}
The effect of impurities on the transport properties of
low-dimensional systems has attracted a great deal of attention
during the last years. In a pioneer work, Kane and Fisher showed
that the transport of a one-channel Luttinger liquid through a
barrier is strikingly different from that in a Fermi-liquid
consisting of non-interacting electrons.\cite{Kane1} Using a
renormalization group technique, they have shown that the behavior
depends crucially on whether the electron-interactions are
repulsive or attractive. For repulsive interactions, $g<1$, the
barrier is relevant, and arbitrary small ones lead to complete
reflection at zero temperature. On the other hand, for attractive
interactions, $g>1$, the barrier is an irrelevant perturbation and
transmission is perfect through even the largest barriers.
Non-interacting electrons ($g=1$) exhibit a behavior intermediate
between these two limits because in this case the barrier is a
marginal perturbation.

Later, the resonant tunneling of a single-channel interacting
electrons through a double barrier was considered and the
time-dependence of the width of the resonances was
discussed.\cite{Kane2,Kane3}
In addition, a Hall bar with two weak barriers was proposed as an
appropriate device for measuring the fractional charge in the
quantum Hall effect.\cite{Chamon2} However, in all those works,
the barriers were considered to be static.


Recently, the transport in a Luttinger liquid in the presence of a
time-dependent impurity was
considered.\cite{Schmeltzer,Sharma,Feldman,Cheng} The model may
describe a constricted quantum wire or a constricted Hall bar at
$\nu = 1/(2n+1)$ with a time-dependent voltage at the
constriction.\cite{Miliken} In particular, Feldman and Gefen
\cite{Feldman} have found an enhancement of the conductance beyond
the value of the quantum of conductance, $G_0 = e^2 / h$, for
strong repulsive electron-electron interactions, $g < 1/2$.

In this paper we study the effect of several {\it time-dependent}
barriers on the transport in an otherwise perfect one-channel
Luttinger liquid. We find that due to the interference term
between the impurities, the conductance can be enhanced {\it also
for weak repulsive interactions}, $1/2<g<1$. In the appropriate
limits, our work reproduces the previous known results for a {\it
single time-dependent} impurity,\cite{Feldman} and for {\it
several static} impurities.\cite{ Chamon2} The model is based on
the bosonization formalism and the calculations are performed
using perturbation theory and the Keldysh technique.

The structure of the paper is the following: in section II we
present our model and derive a general expression for the increase
of the dc current beyond the quantized value $G_0$. We consider
only the backscattering contribution and restrict ourselves to the
lowest order of the perturbation theory, in which, as it is
well-known, forward scattering has no effect on the conductance of
one-dimensional systems.
In section III we compare our results with the ones for a single
impurity,\cite{Feldman} and in section IV we discuss the
multi-impurity case, with an emphasis on the simplest case of a
double-barrier structure, where the novel features are already
present. Our conclusions are drawn in section V.
\section{The Model}

In this section we study the transport properties of a
one-dimensional quantum wire in the presence of several
time-dependent impurities located at positions $x_p$, $p=1,...,N$.
We consider spinless electrons and concentrate on the
zero-temperature case. Using the bosonization technique, the
action of the model reads \cite{Feldman,Chamon,Rao}
\begin{eqnarray}
S &=&\int dt dx
\left\{\frac{1}{8\pi}\left[(\partial_{t}\hat{\Phi})^{2}- v_F^2
(\partial_{x}\hat{\Phi})^{2}\right] \right. \\ \nonumber &-&
\left.\sum_{p}\delta(x-x_{p})W_{p}(t)\left(e^{i\sqrt{g}
\hat{\Phi}(t,x_{p})} e^{i\omega_{0}t}+H.c.\right)\right\},
\end{eqnarray}
where the bosonic field $\Phi$ is related to the charge density
$\hat{\rho}=e\sqrt{g}\partial_{x}\hat{\Phi}/2\pi$, $\omega_0 =
eV/\hbar$ is the Josephson frequency associated with the external
voltage $V$, $W_p(t)$ is an arbitrary pumping function, which
controls the time dependence of the impurity potential, $v_F$ is
the Fermi velocity, and $g$ measures the strength of the
electron-electron interactions.
In the following, we set the Planck constant $\hbar=1$ and the
Fermi velocity $v_F=1$.

We begin our analysis by introducing the backscattering impurity
operator
\begin{equation}
\hat{B}(t)\equiv \sum_{p}W_{p}(t)e^{i\sqrt{g}\hat{\Phi}(t,x_{p})}
e^{i\omega_{0}t}.
\end{equation}
The operator for the backscattering current then reads
\begin{equation}
\hat{I}_{bs}=-i \hat{B}(t)+H.c.
\end{equation}
In order to find the backscattering current at time $t$, we must
evaluate the expectation value
\begin{equation} \langle
\hat{I}_{bs}\rangle=\langle
0|S(-\infty;t)\hat{I}_{bs}S(t;-\infty)|0\rangle, \label{averageI}
\end{equation}
where $\langle 0|$ denotes the initial state, and $S$ is the
scattering matrix. To the lowest order in W,
\begin{eqnarray}  \nonumber
S(t;-\infty) &=& 1- i \int_{-\infty}^{t}dt'[\hat{B}(t') +H.c.], \\
S(-\infty ;t) &=& S^{\ast}(t;-\infty). \label{sinfty}
\end{eqnarray}
Substituting then Eqs.\ (\ref{sinfty}) into Eq.\ (\ref{averageI})
and keeping only the terms of order $W^{2}$ and zero total charge
we obtain
\begin{equation} I_{bs}(t)= \int_{-\infty}^{t}dt'
\langle
0|\hat{B}^{\ast}(t')\hat{B}(t)-\hat{B}(t')\hat{B}^{\ast}(t) + H.c.
|0\rangle,
\end{equation}
where
\begin{equation}\nonumber
\langle 0|\hat{B}^{\ast}(t')\hat{B}(t)|0\rangle =
\sum_{k,j}W_{k}(t')W_{j}(t) e^{i\omega_{0}(t-t')} {\cal J}_{kj},
\end{equation}
\begin{eqnarray}
{\cal J}_{kj}&\equiv& \langle 0|e^{-i\sqrt{g}\hat{\Phi}(t',x_{k})}
e^{i\sqrt{g}\hat{\Phi}(t,x_{j})} |0\rangle\nonumber\\
&=& e^{g\langle
0|\hat{\Phi}(t',x_{k})\hat{\Phi}(t,x_{j})|0\rangle}.
\end{eqnarray}
The Green function of the Bose field reads
\begin{eqnarray} \nonumber
\langle 0|\hat{\Phi}(t',x_{k})\hat{\Phi}(t,x_{j})|0\rangle = &-&
\ln[\delta+i(t'-t+x_{k}-x_{j})] \\ \nonumber &-&
\ln[\delta+i(t'-t+x_{j}-x_{k})],
\end{eqnarray}
where $\delta$ is infinitesimal. Defining now the new variable
$\tau=t-t'$, the backscattering current acquires the form
\begin{eqnarray} \nonumber
I_{bs}(t)=\int_{0}^{\infty}d\tau [e^{i\omega_{0}\tau} - H.c.]
\sum_{k,j} W_{k}(t-\tau) W_{j}(t) \\\nonumber
\times\{[\delta+i(\tau+x_{k}-x_{j})]^{-g}[\delta+i(\tau+x_{j}-x_{k})]^{-g}
-H.c.\}.
\end{eqnarray}
Because $\delta$ has an infinitesimally small value, $ {\rm
Im}\{[(x_{k}-x_{j})^{2}-\tau^{2}]^{-g}\}=0$, for $ |x_{k}-x_{j}|
\geq \tau$. In addition, ${\rm Im}[(-1)^{-g}] = {\rm Im}[e^{-i\pi
g}]=-\sin \pi g$, and we eventually obtain
 \begin{equation}
 I_{bs}(t)=C \sum_{k,j} \int_{|x_{k}-x_{j}|}^{\infty}d\tau
 \frac{\sin(\omega_{0}\tau)  W_{k}(t-\tau)
 W_{j}(t)}{|\tau^{2}-(x_{k}-x_{j})^{2}|^{g}},
\label{Ibs}
\end{equation}
where $C = -4 \sin (\pi g)$. Eq.\ (\ref{Ibs}) is one of the main
results of this paper. It yields in lowest order of perturbation
theory the backscattering current of a Luttinger liquid in the
presence of an arbitrary number of time-dependent impurities, each
of them providing a scattering potential defined by a function $W_j(t)$. 

We can progress further in the analytical calculation of some
specific examples by introducing a simplifying assumption, namely,
that the impurity potential is a periodic function of time,
$W_{j}(t)=W_{j}\cos(\Omega_{j}t+\varphi_{j})$.
In this case, the numerator in the integral in Eq.\ (\ref{Ibs})
reads
\begin{eqnarray} \nonumber
&& \sin(\omega_{0}\tau)W_{k}(t-\tau)W_{j}(t)=
(W_{j}W_{k}/4)\nonumber\\
&\times& \left\{\sin\left[(\omega_{0}-\Omega_{k})\tau+
(\Omega_{k}+\Omega_{j})t+\varphi_{k}+\varphi_{j}\right]  \right. \\
\nonumber &+&
\sin\left[(\omega_{0}+\Omega_{k})\tau-(\Omega_{k}+\Omega_{j})t-\varphi_{k}-
\varphi_{j}\right] \\ \nonumber &+&
\sin\left[(\omega_{0}-\Omega_{k})\tau+(\Omega_{k}-\Omega_{j})t+\varphi_{k}-
\varphi_{j}\right] \\ \nonumber &+& \left.
\sin\left[(\omega_{0}+\Omega_{k})\tau-(\Omega_{k}-\Omega_{j})t-
\varphi_{k}+ \varphi_{j}\right]\right\},
\end{eqnarray}
indicating that in order to obtain a nonzero dc-component of the
current, it is necessary to have at least two modes with the same
frequency. Let us now define
\begin{equation}
F_{g}(\omega,x,\varphi) \equiv \sin (\pi g)
\int_{|x|}^{\infty}d\tau \frac{\sin(\omega
\tau+\varphi)}{|\tau^{2}-x^{2}|^{g}}, \label{Ffunction}
\end{equation}
which can be explicitly evaluated, and reads
\begin{eqnarray} \nonumber
F_{g}(\omega,x,\varphi) &=& \frac{\sin(\pi g)}{2^{3/2-g}}
\left|\frac{\omega}{x}\right|^{g-1/2} \left\{
2\Gamma\left(\frac{1}{2}+g\right)\Gamma(1-2g) \right. \\ \nonumber
&\times& J_{g-1/2}(|\omega x|) {\rm sgn}(\omega)\cos[\varphi -{\rm sgn}(\omega) \pi g]  \\
\nonumber &+&  \left. \Gamma\left(g-\frac{1}{2}\right)
\Gamma(2-2g)J_{1/2-g}(|\omega x|) \sin(\varphi) \right\},
 \end{eqnarray}
where $J_{\alpha}(x)$ are Bessel functions of the first kind.
Assuming that the impurities oscillate with the same frequency,
$\Omega_{k}\equiv \Omega$, we find that for each term $<k,j>$ in
the sum there is a counterterm $<j,k>$ with opposite phase. Hence,
the contribution to the dc current contains terms of the following
form,
$$
\sin\left(\omega\tau+\varphi\right)+\sin\left(\omega\tau-\varphi\right)=
2\sin\left(\omega\tau \right)\cos(\varphi),
$$
yielding,
\begin{eqnarray} \nonumber
I_{dc} &=&-\sum_{k,j}W_{k}W_{j}\cos(\varphi_{j}-\varphi_{k})
\left[F_{g}\left(\omega_{0}+\Omega,x_{k}-x_{j},0\right)
\right. \\
&+& \left. F_{g}\left(\omega_{0}-\Omega,x_{k}-x_{j},0 \right)
\right]. \label{sol}
\end{eqnarray}

Therefore, the dc component of the backscattering current
generated by a set of dynamical impurities oscillating with a
frequency $\Omega$ can be expressed, in first order of
perturbation theory, in terms of the dc current for a set of
static impurities
\begin{equation}
I_{dc}=\frac{1}{2}[I_{st}(\omega_{0}+\Omega)+I_{st}(\omega_{0}-\Omega)],
\label{Idc}
\end{equation}
where the dc current for a set of static impurities reads
\begin{equation} I_{st}(\omega) =
-\sum_{k,j}W_{k}W_{j}\cos(\varphi_{j}-\varphi_{k})
F_{g}\left(\omega,x_{k}-x_{j},0\right).
 \label{static}
\end{equation}
%
The above form of the dc current generalizes the result obtained
by Feldman and Gefen in Ref.\ \onlinecite{Feldman} for the case of
one impurity. Eq.\ (\ref{static}) extends the result obtained by
Chamon {\it et al.} in Ref.\ \onlinecite{Chamon2} for the dc
component of the backscattering current generated by a set of
static impurities. It is worth noting here that the function
$I_{st}(\omega)$ is odd, $I_{st}(-\omega)=-I_{st}(\omega)$, and
can be represented as a product of a dimensional factor and a
function of dimensionless variables $\omega x_{k}$, since
\begin{equation}
F_g(\omega,x,0)\sim|x|^{1-2g}|\omega x|^{g-1/2}J_{g-1/2}(|\omega
x|).
\end{equation}
The dimensionless part remains invariant with scaling
\begin{equation}\label{scaling}
\omega\rightarrow \omega' = \omega \eta, \quad x_{k}\rightarrow
x'_{k}=x_{k}/\eta, \forall k,
\end{equation}
whereas both currents $I_{st}$ and $I_{dc}$ scale with the
dimensional factor
\begin{equation}
I_{st}(\omega,\{x_{k}\})\rightarrow I_{st}(\omega',\{x'_{k}\}) =
\eta^{2g-1} I_{st}(\omega,\{x_{k}\}).
\end{equation}
This leads to a conclusion that the inter-impurity separation can
be fixed without any loss of generality, i.e., the dependence of
the dc current on the inter-impurity separation is the same as on
the Josephson frequency, $\omega_0$, up to a factor, which is a
power of the distance between the impurities. Therefore, by fixing
the distance between the impurities and considering the dependence
of the dc current on the Josephson frequency, we can also obtain
the dependence of the dc current on the distance between the
impurities, as a consequence of the scaling transformation, given
by Eq.\ (\ref{scaling}).


\section{Single time-dependent impurity}

This case has already been discussed in the literature, see Refs.\
\onlinecite{Schmeltzer,Feldman}. Here, we follow the notation of
Refs.\ \onlinecite{Chamon,Feldman}, and calculate the dc
contribution $I_{dc}$ to the total current $I = I_{dc} + I_{ac}$,
for the sake of completeness, as well as to show that our model
correctly reproduces the well-known limit.
\begin{figure}[htb]
\begin{center}
\includegraphics[width=7cm]{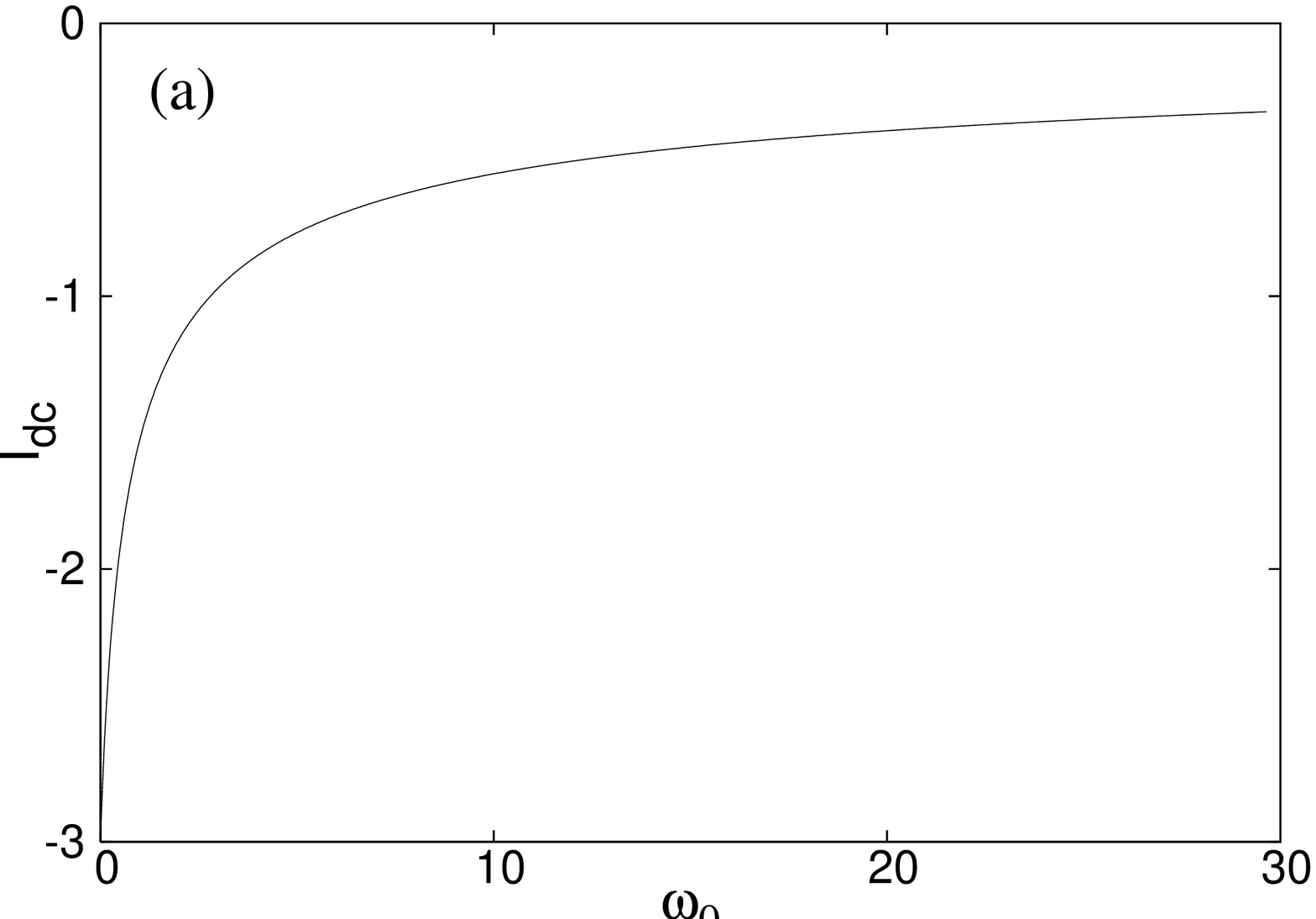}
%
\includegraphics[width=7.4cm]{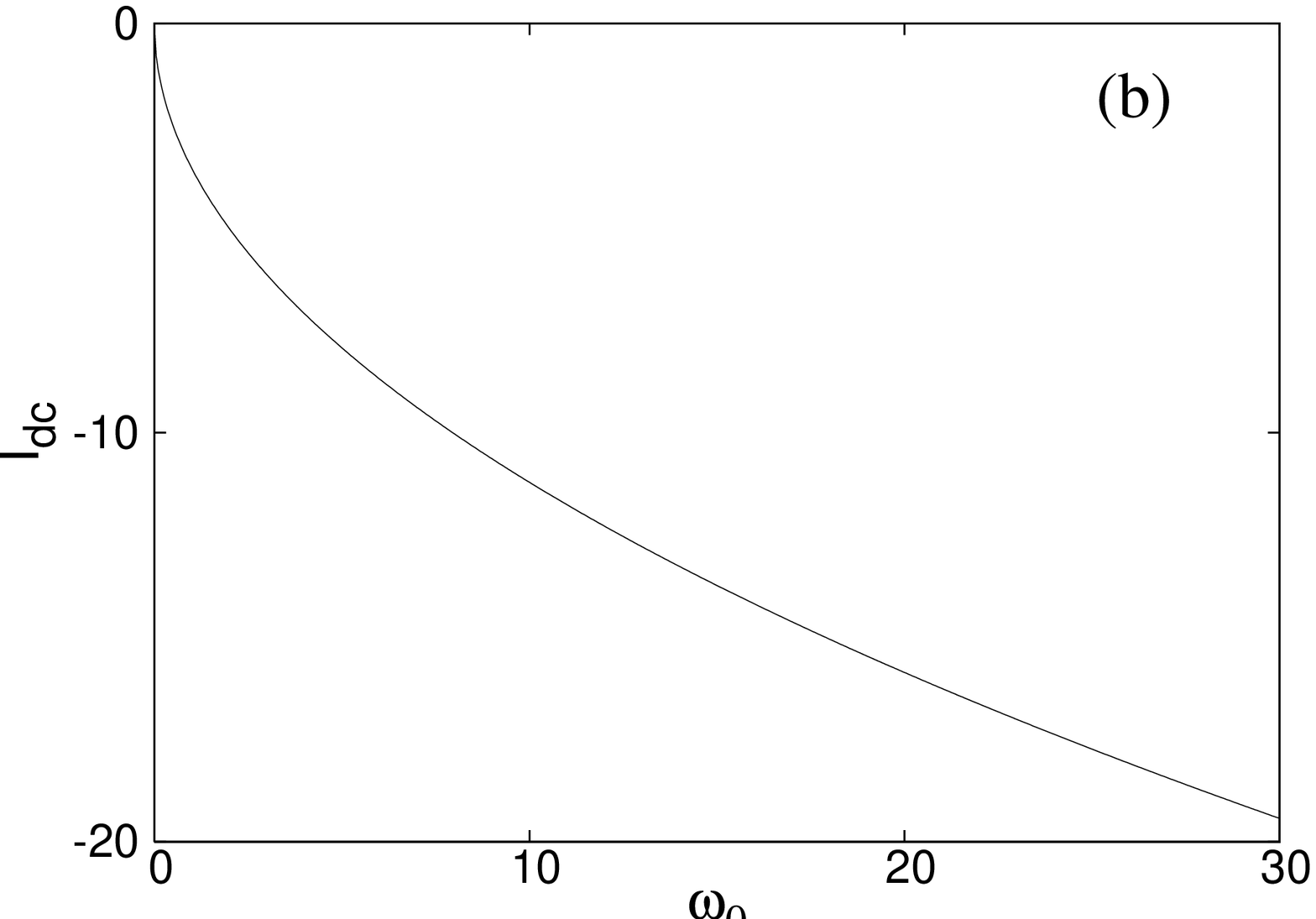}
\end{center}
\caption{\label{fig1}$I_{dc}$ versus $\omega_0$ for one static
impurity with $U=1$. (a) $g = 1/4$. (b) $g = 3/4$. }
\end{figure}
Defining $W_j\equiv U$ and using Eq.\ (\ref{static}), we may
express the dc current for one static impurity as
\begin{equation}
I_{st}(\omega_{0})=-2U^{2}F_{g}\left(\omega_{0},0,0\right),
\end{equation}
where the function $F_{g}\left(\omega_{0},0,0\right)$ can be
determined using asymptotic properties of Bessel functions for
small values of the argument,
\begin{equation}
I_{st}(\omega_0)=-U^2{\rm sgn}(\omega_0)\Gamma(1-2g)\sin (2\pi g)
|\omega_0|^{2g-1}. \label{staticsingle}
\end{equation}
This function is plotted in Fig.\ 1 for different values of the
interaction strength, $g$, and we set $U=1$. Using Eq.\
(\ref{Idc}), the dc component of the backscattering current
generated by the presence of a single time-dependent impurity
reads
\begin{eqnarray} \nonumber
 I_{dc}&=& \frac{U^{2}}{2}\Gamma(1-2g)\sin (2\pi g) [(\Omega -
\omega_{0})^{2g-1} \\
&-&(\Omega + \omega_{0})^{2g-1}], \quad \Omega > \omega_{0},
\label{Idcsingle1}
\\ \nonumber
I_{dc}&=& -\frac{U^{2}}{2}\Gamma(1-2g)\sin (2\pi g) [(\omega_0 -
\Omega)^{2g-1} \\
&+&(\Omega + \omega_{0})^{2g-1}], \quad \Omega < \omega_0,
\label{Idcsingle2}
\end{eqnarray}
which is in agreement with Ref. \onlinecite{Feldman}, after
replacing the dimensionless variables by the dimensional ones.
The correction to the conductance may also be promptly evaluated,
\begin{equation}
\Delta G=\frac{dI_{dc}}{d V}\big|_{V=0}=\frac{e}{\hbar}
\frac{dI_{dc}}{d \omega_{0}}\big|_{\omega_{0}=0},
\end{equation}
which yields
\begin{equation}
\Delta G=\frac{e U^{2}}{\hbar}(1-2g)\Gamma(1-2g)\sin (2\pi
g)\Omega^{2g-2}.
\end{equation}
The interesting result for the time-dependent single-impurity case
\cite{Feldman} is that at strong repulsive interactions, $g <
1/2$, the backscattering current becomes positive for $\Omega >
\omega_0$, thus leading to an increase of the conductance, in
comparison with the impurity-free value, $G_0 = e^2/h$. The ac
contribution to the current may be analogously calculated using
\begin{eqnarray}
I_{ac}=-U^{2}\lim_{d\rightarrow0}\left[F_g\left(\omega_{0}-\Omega,d,2\Omega
t\right)+ F_g\left(\Omega + \omega_{0},d,-2\Omega t\right)
\right],\nonumber
\end{eqnarray}
which yields
\begin{eqnarray} \nonumber
 I_{ac}(t)&=& U^{2}\Gamma(1-2g)\sin (\pi g)\cos(\pi g+2\Omega t) \\
 &\times &[(\Omega - \omega_{0})^{2g-1} -(\Omega + \omega_{0})^{2g-1}], \quad \Omega >
\omega_{0},\nonumber\\ \nonumber
I_{ac}(t)&=&-U^{2}\Gamma(1-2g)\sin
(\pi g)[\cos(\pi
g-2\Omega t)(\omega_0 - \Omega)^{2g-1} \\
&+&\cos(\pi g+2\Omega t)(\Omega + \omega_{0})^{2g-1}], \quad
\Omega < \omega_0. \nonumber
\end{eqnarray}

\begin{figure}[htb]
\begin{center}
\vspace{.5cm}
\includegraphics[width=7cm]{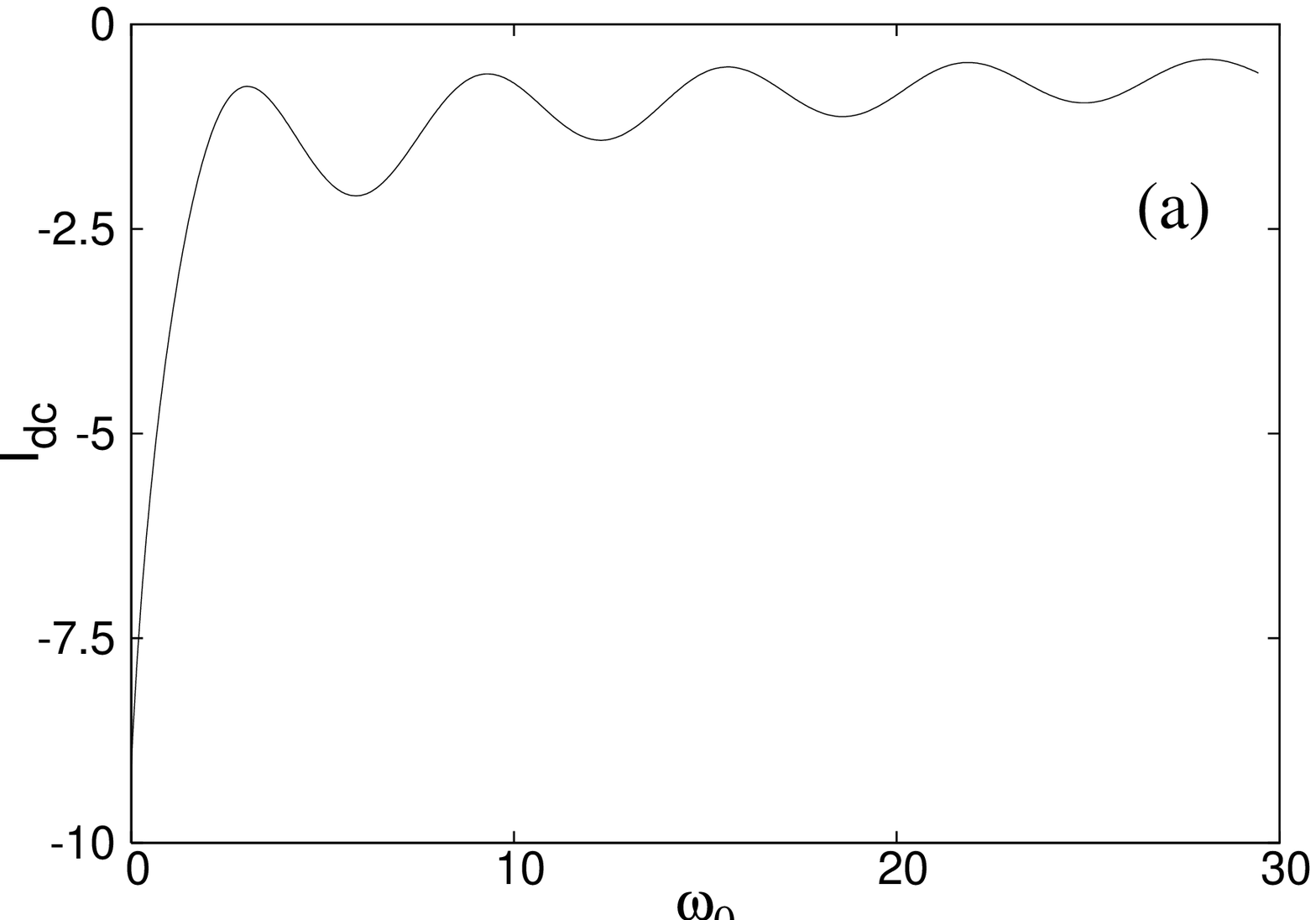}
%
%
%
\includegraphics[width=7cm]{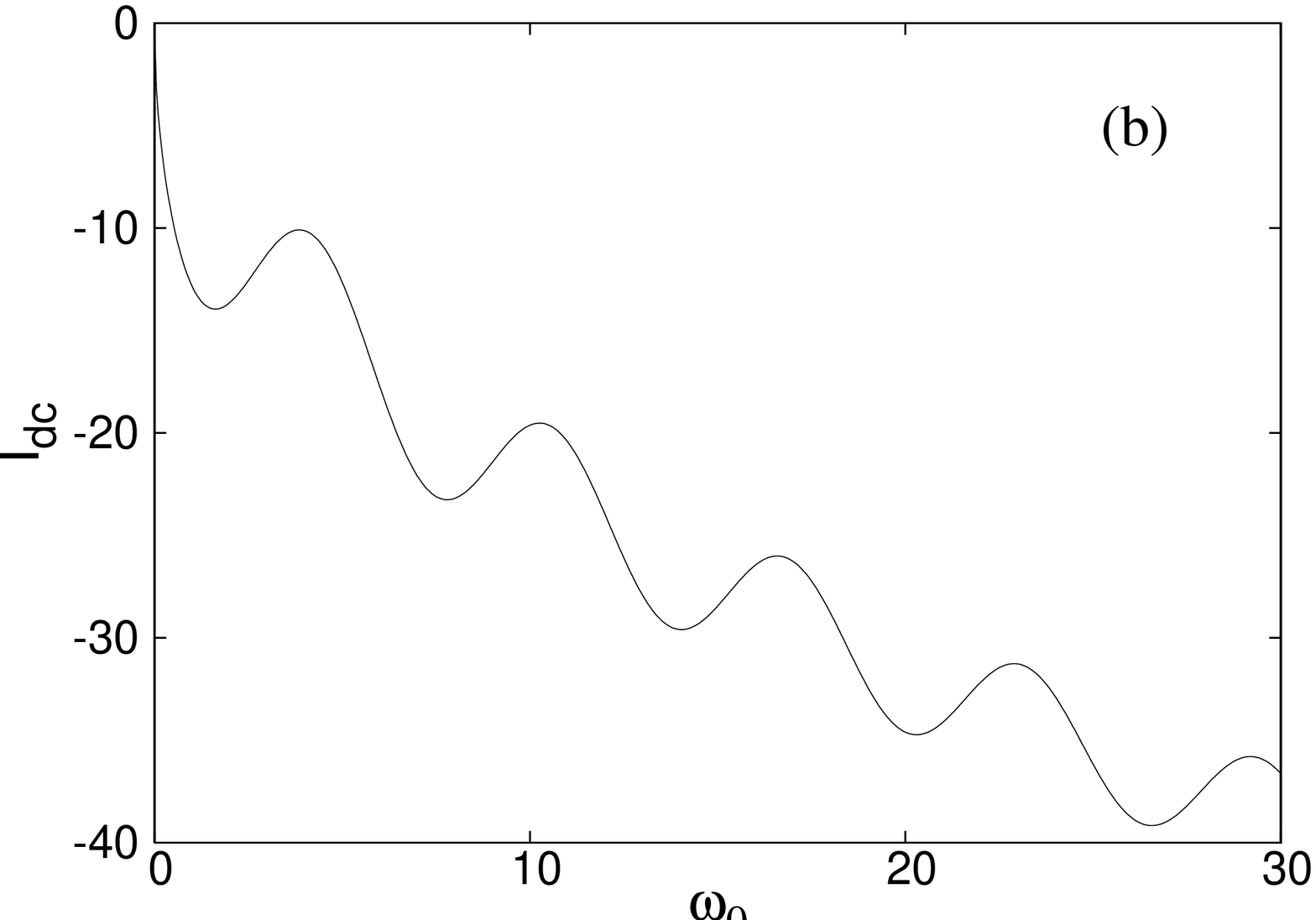}
\end{center}
\caption{\label{fig2}$I_{dc}$ versus $\omega_0$ for two static
impurities with $L=1$ and $U=1$. (a) $g = 1/4$. (b) $g = 3/4$. }
\end{figure}

\section{Several time-dependent impurities on a lattice}

In this section we first concentrate on the simplest case
involving more than one impurity, namely, the case of two
time-dependent impurities separated by the distance $L$. We
consider the case when the two impurities oscillate with the same
frequency, $\Omega$, and allow for a phase shift between them. Let
us first analyze the dc current contribution to the total
backscattering current as a function of the Josephson frequency
$\omega_0$ associated with the external voltage $V$ for two static
impurities (considered before by Chamon {\it et al.} in Ref.\
\onlinecite{ Chamon2} for $g \leq 1/2$). In this case,  Eq.\
(\ref{sol}) yields
\begin{equation}
I_{st}(\omega_{0}) = I_{st}^{(1)}(\omega_{0}) +
I_{st}^{(2)}(\omega_{0}) + I_{st}^{int}(\omega_{0}),
\label{staticdouble}
\end{equation}
where the first two terms are the dc currents for the single
impurity 1 and 2, respectively, given by Eq.\
(\ref{staticsingle}),
and the interference term reads
\begin{equation}\label{staticdoubleint}
I_{st}^{int}(\omega_{0}) = -4W_{1}W_{2} \cos \varphi
F_{g}\left(\omega_{0},L,0 \right).
\end{equation}
After including the functions $F_g$ defined in Eq.\
(\ref{Ffunction}), we obtain the explicit form of the interference
term
\begin{eqnarray} \label{staticdoubleintexpl}
I_{st}^{int}(\omega_{0}) &=& -2W_1 W_2 {\rm {\rm sgn}}(\omega_{0})
\cos (\varphi) L^{1-2g} \\
&\times&\frac{\pi\Gamma(1/2+g)}{2^{1/2-g}\Gamma(2g)} |\omega_{0}
L|^{g-1/2} J_{g-1/2}(|\omega_{0} L|),\nonumber
\end{eqnarray}
where we used the following identity
$$
\sin(2\pi g)  \Gamma(1-2g) \Gamma(2g) = \pi.
$$
For small distances $\omega_0 L\ll 1$ the case of two impurities
reduces to the above-considered single impurity case with an
effective amplitude of the impurity potential $U^{2}=W_1^{2} +
W_2^{2}+2W_1W_2\cos (\varphi)$, which results from the asymptotic
property of the Bessel function
\begin{equation}
\lim_{x\rightarrow 0}|x|^{1/2-g} J_{g-1/2}(|x|)=
\frac{2^{1/2-g}}{\Gamma(1/2+g)}.
\end{equation}

Let us now plot Eq.\ (\ref{staticdouble}) to render the results
more visible. We first consider two impurities with equal
amplitudes of the potential, $W_1 = W_2 = U$, and without the
phase shift, $\varphi = 0$, to simplify the problem. We plot in
Fig.\ 2 the dc current for two static impurities as a function of
$\omega_0$. In Fig.\ 2a, the interaction strength, $g = 1/4$,
whereas in Fig.\ 2b $g = 3/4$. As it was argued above, fixing
$L=1$ does not lead to any loss of generality, since the value of
the current for some other length can be obtained by scaling.
Comparing Fig.\ 1 and Fig.\ 2 we can conclude that the
interference term modifies the dc current in a way that it becomes
a non-monotonous function of the external voltage. The dc current
for the two impurities oscillating with the same frequency
$\Omega$ can now be readily obtained using the result for the
static impurities, Eqs.\ (\ref{staticdouble}) and
(\ref{staticdoubleint}), and Eq.\ (\ref{Idc}), which relates the
current for the dynamical and the static impurities.

The dependence of $I_{dc}$ on $\Omega$ and $\omega_0$ can be
better observed in a 3D plot, see Fig.\ 3a.
\begin{figure}[htb]
\begin{center}
\includegraphics[width=7cm]{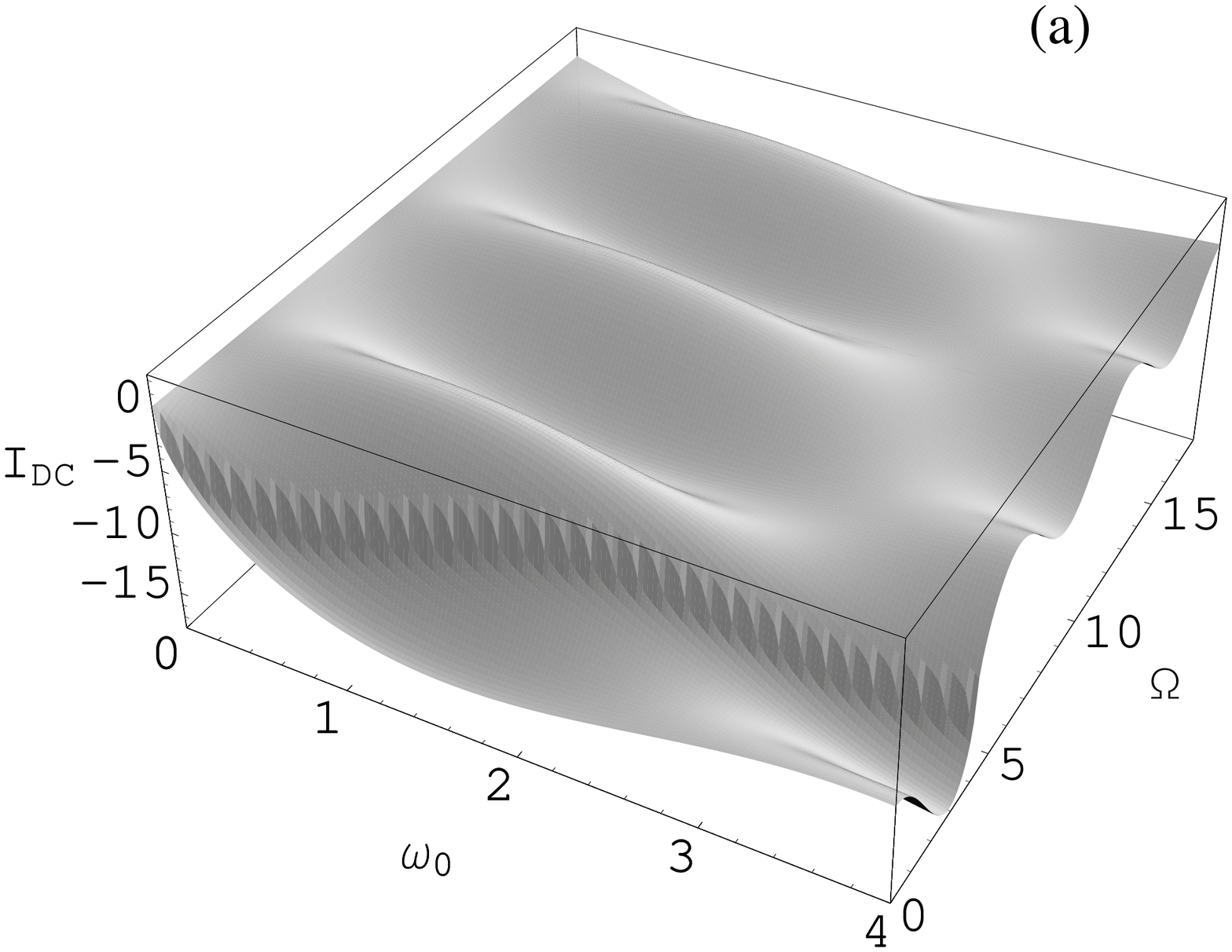}
%
\includegraphics[width=7cm]{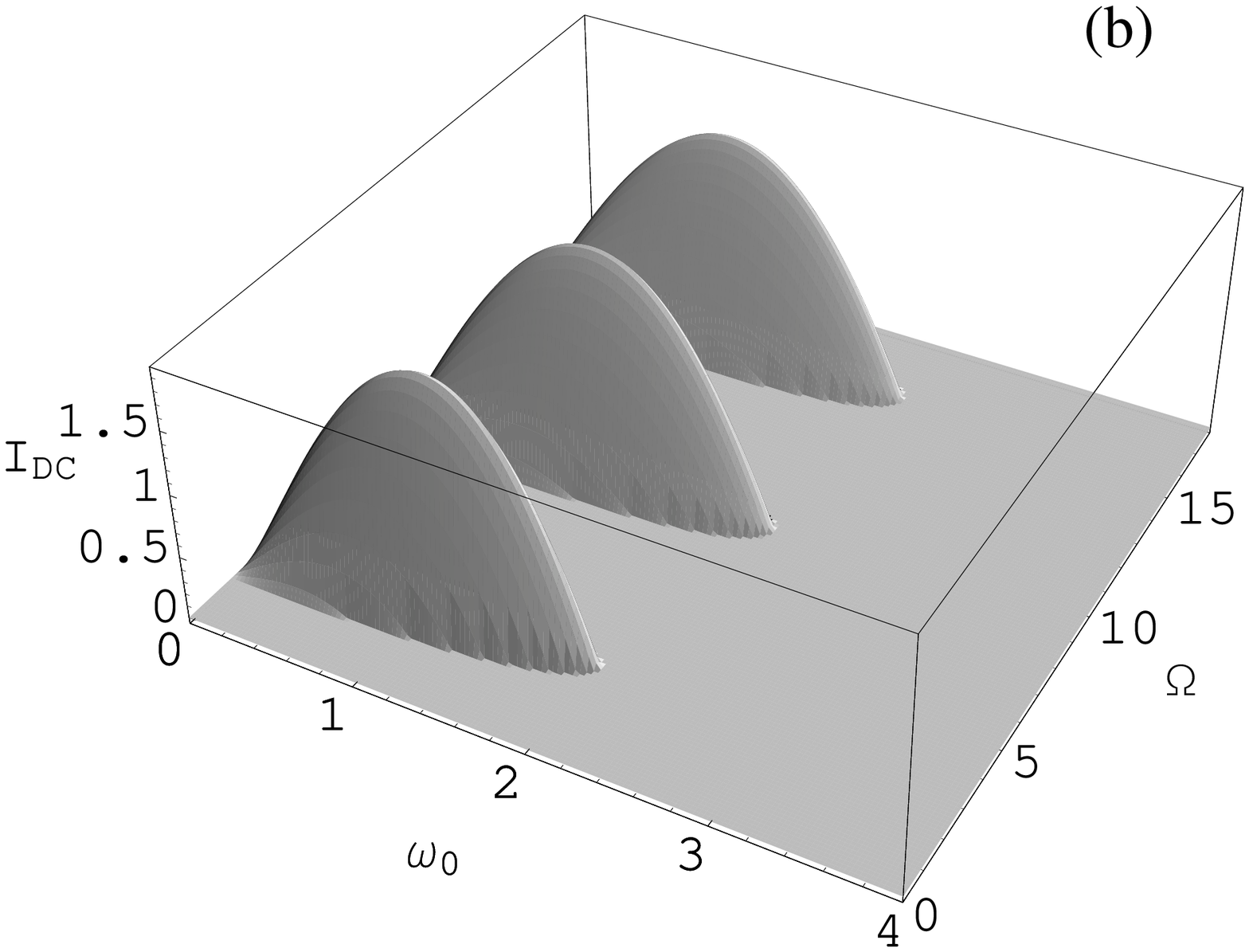}
\end{center}
\caption{\label{fig3} (a) Dependence of the dc current on the
Josephson frequency, $\omega_0$, and the impurity frequency,
$\Omega$, for two impurities with $L=1$, $U=1$, and $g = 3/4$. (b)
the region with a positive dc current, $I_{dc}>0$. }
\end{figure}
In  Fig.\ 3b we selected only the positive contribution in order
to facilitate the visualization of the region where the novel
effects arise. Let us first notice that the backscattering current
enhances the total current {\it even for weaker} ($g> 1/2$)
repulsive interactions, see Fig.\ 3b. The impurity frequency at
which this effect arises coincides with the Josephson frequency at
which the current for the static impurities has a positive slope,
see Fig.\ 2b. This result contrasts with the one obtained for the
single time-dependent impurity, where no positive contribution
exists for $g > 1/2$, since the dc current for a single static
impurity monotonously decreases with the Josephson frequency, see
Fig.\ 4a. On the other hand, the presence of the interference term
in the case of two impurities gives rise to a more complicated
non-monotonous behavior of the current with the external voltage,
which leads to an enhancement of the total current even for weak
repulsive interactions.
\begin{figure}[htb]
\begin{center}
\includegraphics[width=7cm]{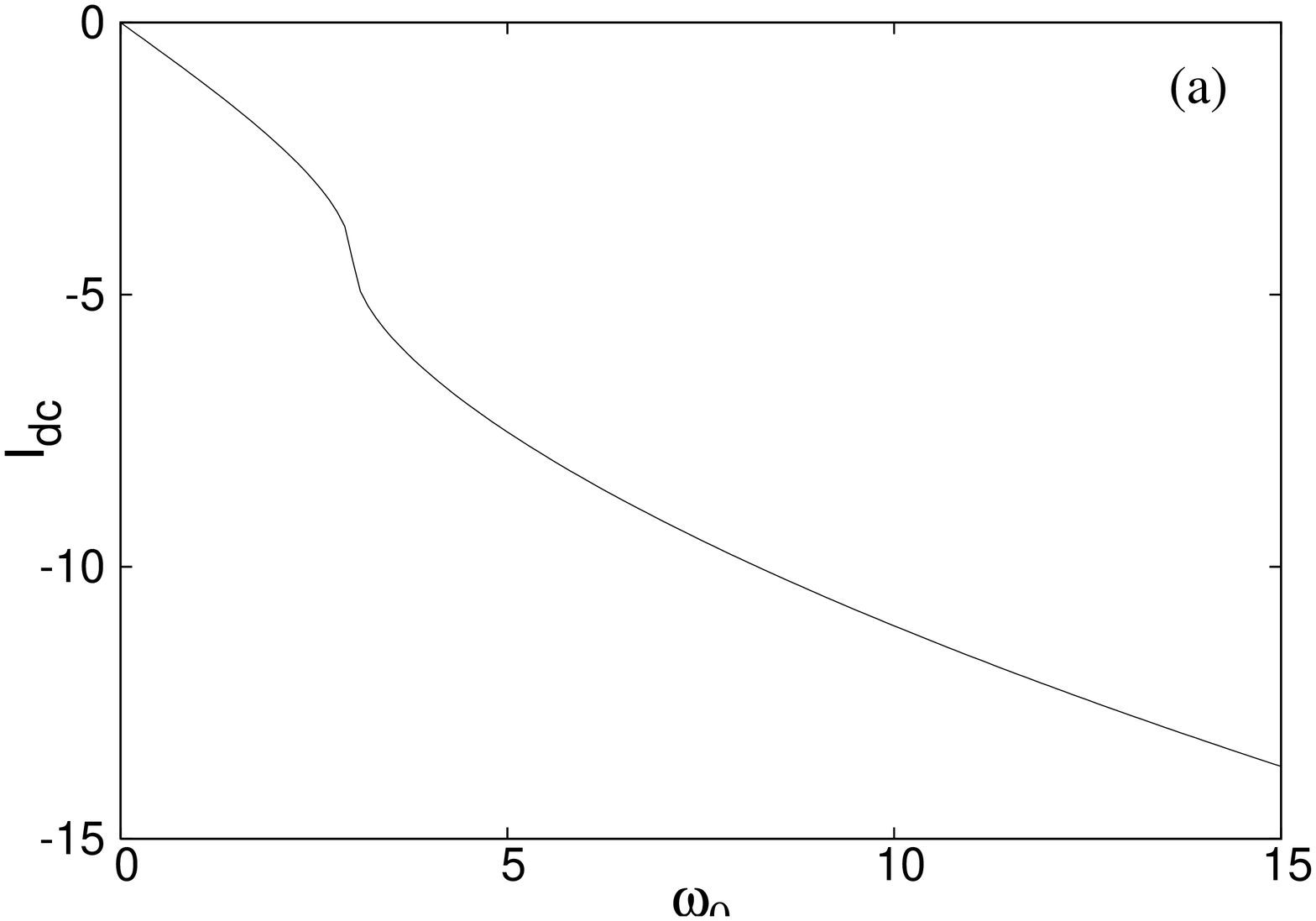}
\includegraphics[width=7cm]{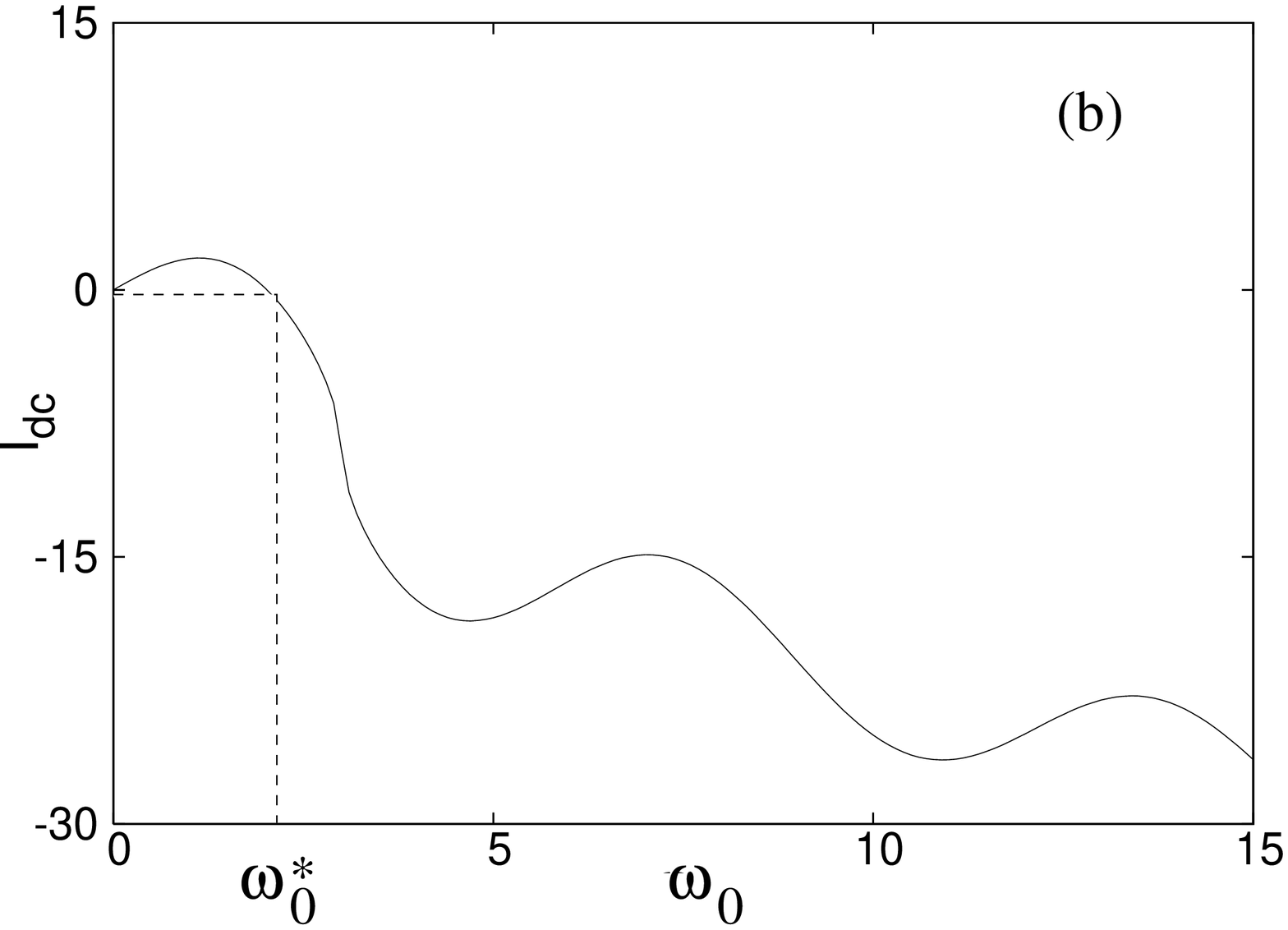}
\end{center}
\caption{\label{fig4}$I_{dc}$ versus $\omega_0$ for $g = 3/4$,
$\Omega=3$, and $U=1$. (a) one impurity. (b) two impurities with
$L=1$. }
\end{figure}

For illustrative purposes we consider a particular section of the
3D plot in Fig.\ 3a defined by the plane $\Omega = 3$ for the
novel case where there is an increase of the current, namely, $g =
3/4$.
The dc current dependence on the Josephson frequency, shown in
Fig.\ 4b, indicates that the total current is increased for small
values of $\omega_0$. Beyond a critical value $\omega_0^*$ the
current becomes negative, and at $\omega_0 = \Omega$ a step can be
observed.

This effect can be understood by considering the relationship
between the dc currents for the time-dependent and the static
impurities. Recalling that $I_{st}(\omega)$ is an odd function, we
may rewrite Eq.\ (\ref{Idc}) as
\begin{equation} I_{dc}=
\frac{1}{2}[I_{st}(\Omega+\omega_{0})-I_{st}(\Omega-\omega_{0})].
\label{Idc2}
\end{equation}
For a positive dc-component of the backscattering current,
$I_{dc}>0$, the previous equation implies
\begin{equation}\label{ineq1}
I_{st}(\omega_{0}+\Omega)>I_{st}(\Omega-\omega_{0}).
\end{equation}
The odd function $I_{st}(\omega)$ is negative for positive
arguments, the inequality can thus only hold if
$\Omega-\omega_{0}>0$, since for $\Omega-\omega_0<0$, the function
$I_{st}$ is positive, $I_{st}(\Omega-\omega_0<0)>0$.
In the one-impurity case, the function $I_{st}(\omega)$
monotonously increases with $\omega$ for $g<1/2$, hence the
inequality (\ref{ineq1}) holds for $\Omega > \omega_0$, resulting
in a positive dc current. On the other hand, for weak repulsive
interactions, $g>1/2$, the current $I_{st}$ is a decreasing
function of its argument, the inequality (\ref{ineq1}) is not
obeyed, and, consequently, the dc-component of the backscattering
current is negative. In the case of two impurities, the function
$I_{st}$ is non-monotonous even for weak repulsive interactions,
$g>1/2$, meaning that there are intervals of frequency in which
the function $I_{st}$ increases. If the frequency of the impurity
belongs to one of such intervals, according to Eq.\ (\ref{ineq1}),
the total current is increased, at least for a small Josephson
frequency, $\omega_0\ll\Omega$. This effect arises as a
consequence of the interference term, given by Eq.\
(\ref{staticdoubleintexpl}), which results from the spatial
correlations in the Luttinger liquid that influence the
backscattering current due to the presence of two point-like
impurities.

In general, for any number of weak impurities, the dc-component of
the backscattering current is positive, if for some frequencies
$0<\omega_{1}<\omega_{2}$, the static current obeys
\begin{equation}\label{posIdc}
I_{st}(\omega_{1})<I_{st}(\omega_{2}).
\end{equation}
The corresponding positive value of the dc current, according to
Eq.\ (\ref{Idc}), reads
\begin{equation}
I_{dc}=\frac{1}{2}[I_{st}(\omega_{2})-I_{st}(\omega_{1})]=
\frac{1}{2}[|I_{st}(\omega_{1})|-|I_{st}(\omega_{2})|],
\end{equation}
where the frequencies $\omega_1$ and $\omega_2$ are related to the
Josephson frequency, $\omega_0$, and the frequency of the impurity
potential, $\Omega$, as follows
\begin{equation}\label{Omegaandomega0}
\Omega=\frac{1}{2}(\omega_{2}+\omega_{1}), \quad
\omega_0=\frac{1}{2}(\omega_{2}-\omega_{1}).
\end{equation}
The criterium for a positive dc-component of the backscattering
current, given by Eq.\ (\ref{posIdc}), implies that there exists
an interval of frequencies where the static current increases.
Then for the impurity frequency, $\Omega$, belonging to such an
interval, and the Josephson frequency $\omega_0\ll\Omega$, using
Taylor expansion of the static current, $I_{st}$, in Eq.\
(\ref{Idc2}) and keeping only the linear term in Josephson
frequency, we obtain
\begin{equation}
I_{dc}=I'_{st}(\Omega)\omega_{0}+\mathcal{O}(\omega_{0}^2).
\end{equation}
Therefore, in the case of weak impurities when the expansion to
the lowest order of perturbation theory is valid, the
backscattering current is positive for any frequency of the
impurity potential, $\Omega$, such that the static current has a
positive slope, $I'_{st}(\Omega)>0$, and the Josephson frequencies
$\omega_0\ll\Omega$. However, this is not the most general
situation, since it may occur, depending on the form of the static
current, that for some frequency $\Omega$, the static current
decreases, $I'_{st}(\Omega)<0$, but the frequency $\omega_0$ is
such that the inequality (\ref{ineq1}) is still obeyed.

Let us now return to the case of two impurities, where, as we will
show below, the increase of the total current occurs only for the
frequencies of the impurity potential at which the static current
increases. The static current in this case is such that for two
consecutive local minima and local maxima,
$\omega_{min1}<\omega_{max1}<\omega_{min2}<\omega_{max2}$, it
obeys
$I_{st}(\omega_{max1})>I_{st}(\omega_{min1})>I_{st}(\omega_{max2})>I_{st}(\omega_{min2})$,
see Fig.\ 2b. Then, the maximal current enhancement reads
\begin{equation}
I_{dc}=\frac{1}{2}[I_{st}(\omega_{max1})-I_{st}(\omega_{min1})].
\end{equation}
The corresponding frequency of the impurity potential and the
Josephson frequency are given by Eq.\ (\ref{Omegaandomega0}) with
$\omega_1\equiv\omega_{min1}$ and $\omega_2\equiv\omega_{max1}$.
%
%
%
In the case of two impurities, the current enhancement $I_{dc}>0$
occurs only for the impurity frequencies $\Omega$ such that
$I'_{st}(\Omega)>0$, since the static current decreases faster
than it increases. Moreover, due to the quasi-periodicity of the
function $I_{st}$, each interval of frequencies where
$I'_{st}(\Omega)>0$ is followed by an interval with
$I'_{st}(\Omega)<0$ in which the current enhancement never occurs.
Therefore, the set of impurity frequencies such that the
dc-component of the backscattering current is positive has a
band-like structure. Finally, for a fixed impurity frequency, the
maximum value of the Josephson frequency, $\omega_0^*$, for which
is possible to observe the current enhancement is limited to the
maximal $\omega_0$ such that
$I_{st}(\Omega-\omega_0)=I_{st}(\Omega+\omega_0)$.

\begin{figure}[htb]
\begin{center}
\includegraphics[width=7cm]{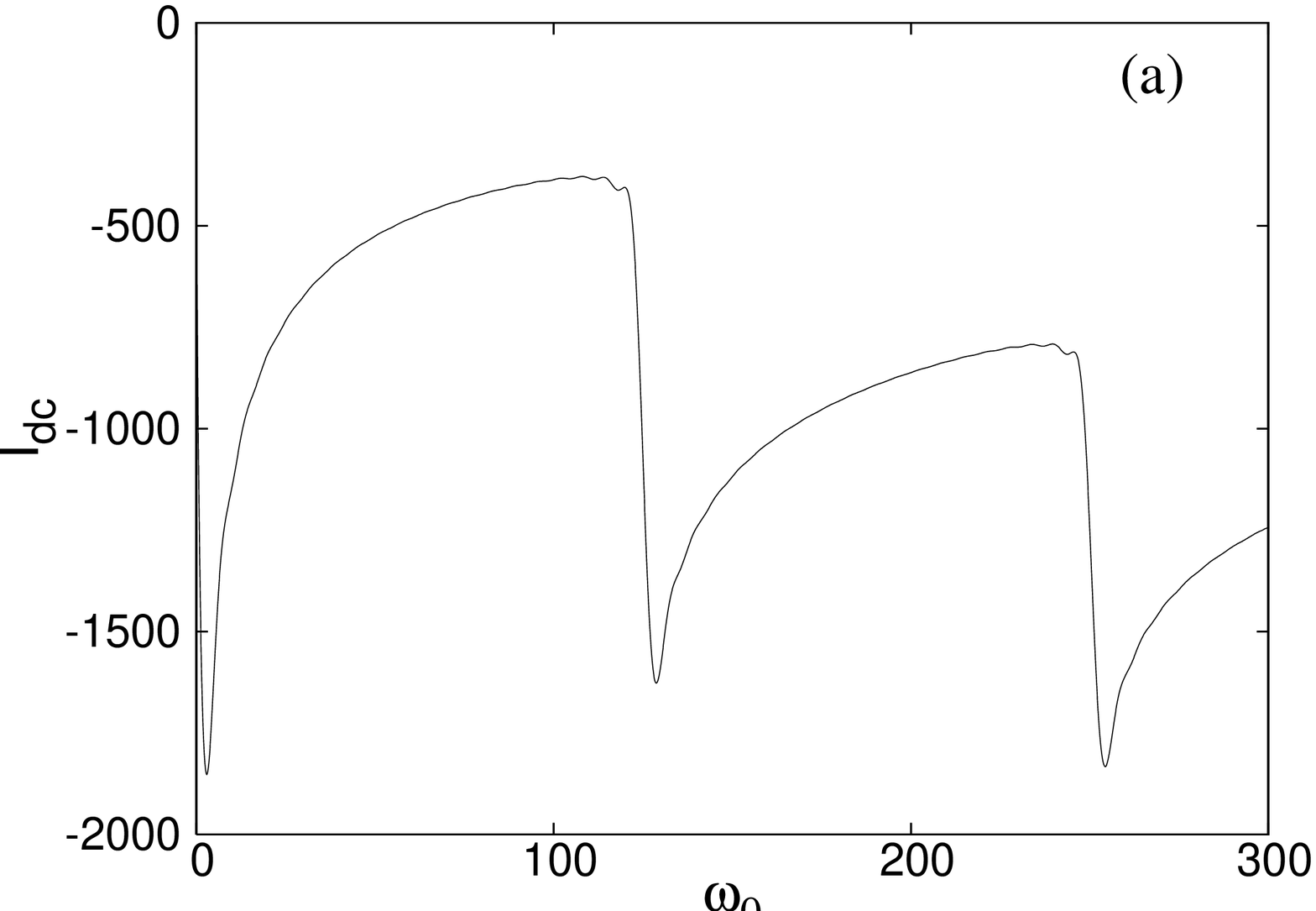}
%
\includegraphics[width=7cm]{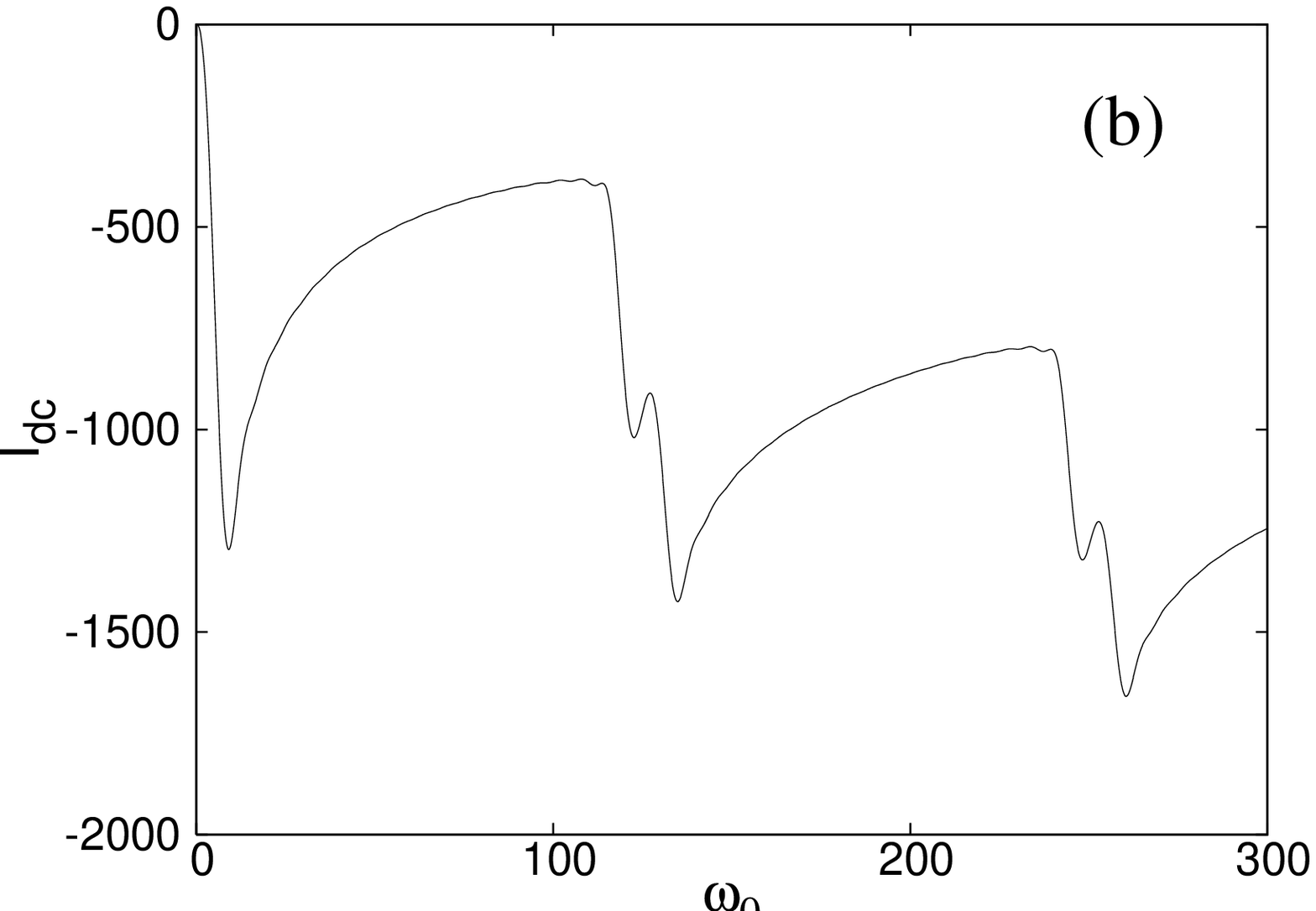}
\end{center}
\caption{\label{fig5}$I_{dc}$ versus $\omega_0$ for $N=20$ static
impurities homogenously distributed on a lattice with the length
$L=1$ and with the phase shift $\Delta\varphi$, for $g=3/4$ and
$U=1$. (a) $\Delta\varphi = 0$. (b) $\Delta\varphi = \pi/10$.}
\end{figure}

We now discuss the dc current for $N$ time-dependent impurities
homogeneously distributed on a lattice, and separated from each
other by the distance $d=L/(N-1)$, in the case of weak repulsive
interactions, $g>1/2$. Since the dc current for the dynamical
impurities is related to the current for the static ones, Eq.\
(\ref{Idc}), in the following we only consider the latter. In this
case, the current consists of $N$ single-impurity contributions
and an additional term, which is a sum of the interference
currents over all possible pairs of impurities. The interference
term of a pair of impurities, located at $x_j$ and $x_k$, given by
Eq.\ (\ref{staticdoubleint}), is quasi-periodic in frequency with
period $2\pi/|x_j-x_k|$. As we discussed above, the values of the
impurity frequency where an increase of the dc-current occurs form
a band-like structure, and thus the values of $\Omega$ for which
$I_{dc}>0$ in the case of $N$ impurities are determined from the
overlap of the bands for each pair of impurities. This overlap
occurs within the frequency range determined by the widest band
formed by the pairs of impurities, which is the one for the two
nearest-neighboring impurities, and thus has the period $2\pi/d$.
The frequency dependence of the dc current is displayed in Fig.\
5a for $N=20$ impurities homogeneously distributed on a lattice
with the length $L=1$. It shows that the frequency region with
positive dc current is approximately 20 times larger than in the
case of two impurities, in agreement with the above arguments.
%
Finally, in Fig.\ 5b we observe that a finite phase shift between
the impurities leads to the appearance of a new region where the
function $I_{st}(\omega_0)$ increases, and thus generates new
bands with a positive dc component of the backscattering current.

%
%
%
%
%
%
%

\section{Conclusions}

In this work, we studied the transport properties of a Luttinger
liquid in the presence of several weak time-dependent point-like
impurities. Our approach is based on the bosonized version of the
Luttinger liquid Hamiltonian, in which we considered the current
of the electrons backscattered by the impurities in lowest order
of perturbation theory. We first found the backscattering current
induced by a set of weak impurities with an arbitrary
time-dependent potential, and we then concentrated on the
particular case of a periodic potential. The obtained form of the
dc current for several impurities oscillating with the same
frequency, given by Eq.\ (\ref{sol}), generalizes the result for a
single dynamical impurity obtained by Feldman and Geffen in Ref.\
\onlinecite{Feldman}. Moreover, we showed that the dc current
generated by a set of time-dependent impurities with a frequency
$\Omega$ in the presence of an external voltage
$\omega_0=eV/\hbar$ reduces to the sum of the currents for static
impurities subjected to an external voltage $|\omega_0\pm\Omega|$,
as described by Eq.\ (\ref{Idc}). This result enabled us to reduce
the problem of calculating the dc current of the dynamical
impurities to the one of the static impurities.

We applied the result for the backscattering current to a single
time-dependent point-like impurity, and showed that it reproduces
the result obtained by Feldman and Gefen, i.e., that the current
enhancement is only possible for strong repulsive interactions,
$g<1/2$. We then concentrated on two dynamical impurities. In this
case, the presence of spatial correlations in the Luttinger liquid
leads to a non-monotonous behavior of the static current, which,
in turn, gives rise to a positive dc-component of the
backscattering current {\it even for weak repulsive interactions},
$g>1/2$. Using the form of the dc current for a set of dynamical
impurities in terms of the corresponding static currents, given by
Eq.\ (\ref{Idc}), we found that for two dynamical impurities the
current enhancement occurs only for the values of the impurity
frequency for which the static current (\ref{static}) has a
positive slope, $I'_{st}(\Omega)>0$. Due to the quasi-periodic
form of the static current for two impurities, the values of the
impurity frequency for which the total current is enhanced form a
band-like structure. Finally, we considered several impurities
homogeneously distributed on a lattice, and found that an increase
of the total current also occurs for weak repulsive interactions,
$g>1/2$. The frequencies of the impurity potential at which the
increase takes place have a band-like structure, as in the case of
two impurities, but the width of the band scales with the number
of impurities, for a fixed length of the lattice.

%
%
%

\end{document}